# The Exceptionally Soft X-ray Spectrum
# of the Low-mass Starburst Galaxy NGC 1705

G. Hensler and R. Dickow

*Institut für Astronomie und Astrophysik, Universität Kiel, Olshausenstr. 40, D-24098 Kiel, Germany*

N. Junkes

*Astrophysikalisches Institut Potsdam, An der Sternwarte 16, D-14482 Potsdam, Germany*

J.S. Gallagher

*Department of Astronomy, University of Wisconsin-Madison, 5534 Sterling Hall, 475 N. Charter St., Madison, WI 53706-1582, USA*

## ABSTRACT

NGC 1705 is one of the optically brightest and best studied dwarf galaxies. It appears to be in the late stage of a major starburst and contains a young super star cluster. Type II supernovae are therefore likely to have been a major effect in the recent evolution of this galaxy and are likely to have produced a superbubble whose affects on the low-density ambient interstellar medium can be ideally studied.

ROSAT PSPC observations of this galaxy reveal two striking blobs of X-ray emission embedded in Hα loops which can be interpreted as both sides of the upper plumes of the same superbubble. These sources are a surprise. They are much softer than those observed from other starburst dwarf galaxies, and are so soft that they should have been blocked if the observed Galactic HI column density were uniformly distributed across NGC 1705 or if the sources were embedded in the HI disk of NGC 1705. In addition, the total X-ray luminosity in the ROSAT energy band of $1.2 \times 10^{38}$ erg s$^{-1}$ is low in comparison to similar objects.

We discuss possible models for the two X-ray peaks in NGC 1705 and find that the sources most likely originate from relatively cool gas of one single superbubble in NGC 1705. The implications of the exceptional softness of these sources are addressed in terms of intrinsic properties of NGC 1705 and the nature of the foreground Galactic absorption.

*Subject headings:* Galaxies: active – Galaxies: compact – Galaxies: individual: NGC 1705 – Galaxies: irregular – Galaxies: starburst – X-rays: galaxies

## 1. Introduction

NGC 1705 is a nearby low-mass irregular galaxy that shows striking signatures in Hα of strong ongoing star formation activity. It was classified by Sandage & Brocato (1979) as an amorphous galaxy. Meurer et al. (1992, hereafter MFDC) as well as Melnick et al. (1985) emphasized the unusual composite structure of NGC 1705. It contains several compact star clusters around a luminous central 'super star cluster' (SSC). These less luminous blue star clusters provide the ionization for the resolved high-surface brightness HII regions and represent the central star-formation activity of a high-surface brightness population which has been formed from about 1 Gyr in the past until the present (MFDC, Quillen et al. 1995). These characteristics are consistent with the selection criteria by Thuan & Martin (1981) for blue compact dwarf galaxies (BCDGs). Finally, a low-surface brightness underlying stellar population has an estimated age of 3 Gyrs.

The recent starburst episode is dominated by the primary central SSC, NGC 1705-1 (Melnick et al. 1985; O'Connell et al. 1994), with an age of about 13-15 Myr. Meurer et al. (1995) reported another cluster, NGC 1705-2, that only appears to overlap with NGC 1705-1 but is more extended and probably separated (Meurer, priv. comm.). Recently, Ho & Filippenko (1996) determined the velocity dispersion of stars in NGC 1705-1 from HIRES echelle spectra taken with the Keck telescope. They derived a virial mass of $(8.2\pm2.1)\times10^4$ $M_\odot$ and predict common galactic globular cluster properties for NGC 1705-1 after 10-15 Gyr. Heckman & Leitherer (1997, hereafter HL97) used UV spectra from the HST to conclude that the presently most massive star in the NGC 1705-1 lies between 10 and 30 $M_\odot$ because of the absence of stellar wind profiles; they deduced an SSC age of $10^{+10}_{-3}$ Myr.

SSCs are a recently more commonly detected phenomenon in actively star-forming dwarf as well as in normal galaxies (Ho 1997). For comparison, the starburst dIrr galaxy NGC 1569 contains two SSCs, for which a clear connection with X-ray bubbles is established from ROSAT HRI observations (Heckman et al. 1995).

The distance deduced by O'Connell et al. (1994) of almost 5 Mpc was used for our analysis (whereas HL97 adopted 6.2 Mpc). The gas mass of $M_{HI}$ was determined by MFDC as $8.7\times10^7 M_\odot$ (for their distance of 4.7 Mpc) and for $M_{HII}$ we adopted $10^7 M_\odot$.

NGC 1705 was observed with IUE in the UV by Lamb et al. (1985) who noted it as being in a "mild star-formation burst". In addition, the Hα image of NGC 1705 shows giant loops and arcs that are strikingly suggestive of explosive events and are similar to those of other starburst dwarf galaxies (SBDGs), like NGC 5253 or NGC 1569.

In the optical, characteristics are typical for blue irregular galaxies (Hunter et al. 1989). The spectra by MFDC detected split emission lines with typical radial velocity differences of about 100 km s$^{-1}$ over much of the central region of the galaxy. This agrees with the signature of outflowing gas recently seen in UV absorption lines (Sahu & Blades 1997). A reasonable model for NGC 1705 is for the outflowing gas to be expelled by the central SSC during the past $10^7$ yrs (Suchkov et al. 1994). Filamentary structures of ionized gas are a prominent feature of NGC 1705. Similar ionized structures have also been found in other actively star-forming small galaxies and in some cases reflect the production of superbubbles by OB stars (e.g. Hunter & Gallagher 1990,1997; Marlowe et al. 1995).

Since NGC 1705 contains one dominant SSC that is about 13 Myr old, type II supernova (SNII) explosions are expected to have accumulated and formed one coherent superbubble. Because of the low gravitational potential, the hot gas should expand, sweep up ambient interstellar gas and possibly escape from the galaxy (Marlowe et al. 1995). The expansion was already found by MFDC due to the radial velocities of the Hα filaments and was recently confirmed from interstellar UV absorption lines (HL97, Sahu & Blades 1997). The latter authors could also distinguish between lines originating in NGC 1705 and Galactic ones.

In this paper, we report the discovery of exceptionally soft X-ray emission from NGC 1705 and describe the basic circumstances which allow us to observe this source. We will discuss the state of the hot gas phase and the expanding superbubble in NGC 1705 in a second paper (Hensler et al. 1998, hereafter Paper II).

## 2. X-ray observations

NGC 1705 has been observed with the ROSAT PSPC in 1994 for 22487 s. From the galaxy itself, we collected only 138 counts. Fig.1 presents a small cut out of the PSPC field as an overlay of the X-ray con-tours with the Hα image taken from Meurer (1989). The position of the ROSAT image is adapted so that the G5 star CPD-53 769 coincides with a lower X-ray maximum. This requires that the original PSPC coordinates be corrected by 8" to the northeast (almost perpendicular to the connection of both bubbles). Furthermore, the brightest X-ray source to the W coincides with a star and we fit the optical spot with the maximum contour.

The background flux amounts to $(8.6\pm3.1)\times10^{-4}$ cts arcmin$^{-2}$ s$^{-1}$ and contributes up to 37% to the total X-ray flux. This is determined from three selected circular fields outside of NGC 1705 (two to the S, one to the NE) that are empty of any X-ray point source and cover a total area of 4.4 arcmin$^2$ (for details and also the whole PSPC field, see Paper II). We detect four X-ray maxima in the field of NGC 1705: The strongest of these sources provides almost 80% of the total background-subtracted X-ray flux in the field and is located 106" westward from the center of the optical body of NGC 1705. It falls outside the area covered in fig.1, but can be perceived from the increasing contours at the right-hand edge. Its connection with NGC 1705 is reasonably excluded; optically it seems to be a foreground object with a power-law spectrum, although the source is not yet identified. Another maximum is located 87" to the NE of NGC 1705 at a projected distance of 2.1 kpc, and seems also to be disconnected from the Hα body. Nevertheless, the H I distribution (Meurer 1994) shows that this maximum is located at the eastern edge of the H I disk so that an embedded X-ray source would be plausible. However, its hardness ratio of +0.72 excludes a normal superbubble but leads to the assumption that one is dealing with an active extragalactic background source (Paper II).

Because the two X-ray maxima located just to the east and west of the center of NGC 1705 lie within the region covered by Hα emission, these sources likely belong to NGC 1705. They both appear to be surrounded by the ionized loops; their structures are very suggestive of hot X-ray bubbles surrounded by cool gas shells.

A comparison with the H I disk of NGC 1705 (Meurer 1994) further shows that the line connecting both maxima is oriented almost perpendicular to the H I plane. The X-ray sources are symmetrically displaced by about 30" (0.7 kpc) from the center. Because of this symmetry and the existence of only one central SSC, it is likely that we are seeing a single hot superbubble produced by accumulated SNII explosions in the SSC. In this picture, the main body of the superbubble is embedded within the H I disk, and therefore hidden in X-rays, while the two observed X-ray sources represent those regions that have expanded high above the H I disk plane, where absorption declines.

The X-ray spectra of both NGC 1705 sources are similar and abnormally soft, as shown in fig.2. In contrast to other SBDGs and blue irregular galaxies that have been detected as X-ray sources, e.g. NGC 1569 (Heckman et al. 1995), NGC 4449 (Bomans et al. 1997, Dickow 1995), NGC 5253 (Martin & Kennicutt 1995, Dickow 1995), or He 2-10 (Hensler et al. 1997) the spectrum of NGC 1705 has its maximum at 0.2 keV and declines steeply to higher energies, so that the flux approaches zero at energies larger than 0.7 keV. All of the above SBDGs are still bright between 0.7-1.0 keV, and even possess maxima there corresponding to the peak of the PSPC sensitivity. Only NGC 4449 exhibits also a (second) maximum between 0.2-0.3 keV in the integrated flux (Dickow 1995) which has also been determined as the "very soft" thermal component by Della Ceca et al. (1997) with kT=0.25 keV. This soft X-ray component stems from two (out of five) X-ray maxima (#4 and #5 in Bomans et al. (1997)) within NGC 4449. Nevertheless, only the X-ray source #5 which is a diffuse area in the south of NGC 4449 shows an extremely soft spectrum with the X-ray flux vanishing at energies above 0.9 keV. However, no other SBDG is known to have *only a soft X-ray component.*

The detection of such soft X-ray emission is mysterious. Even a small column density of $N_{HI}$ within NGC 1705 that would add to the Galactic foreground of $N_{HI,GAL}=3.5\times10^{20}$ cm$^{-2}$ (Dickey & Lockman 1990) should block the soft X-radiation. Otherwise, we would have to deal with an extremely luminous soft X-ray and EUV radiator which would have to show up in EUVE or ROSAT WFC surveys. This is not the case because the EUVE observes an upper-limit count rate e.g. in the alc filter between 156-234Å of almost 0.005 cts/s/keV.

Since the observed count rate declines steeply already below 0.5 keV, only a narrow spectral range is available to determine the type of the spectrum with an inherently large uncertainty (Paper II). A Raymond-Smith spectrum with cosmic abundances provides only a reasonable fit for an $N_{HI}$ value much below the Dickey-Lockman value according to the ex-treme softness of both spectra. E.M. Arnal (1995) obtained by a recent measurement a lower Galactic $N_{HI,GAL}$ of $1.35\times10^{20}$ cm$^{-2}$ with a higher resolution of almost 30" toward NGC 1705. HL97 also find a lower $N_{HI,GAL}=2.0\times10^{20}$ cm$^{-2}$ than Dickey & Lockman from the red wings of Ly$\alpha$ absorption lines. However, even with these lower Galactic column densities, the very soft X-ray spectrum is difficult to explain. NGC 1705 appears to have the softest X-ray spectrum of any starburst galaxy so far detected with ROSAT (Trinchieri, private communication). With $L_X$ of only $1.26\times10^{38}$ erg s$^{-1}$ for the low $N_{HI}$ value, NGC 1705 also has the faintest X-ray luminosity of any X-ray detected SBDG. For comparison, $L_X$ in the ROSAT energy band (0.1-2.4 keV) amounts to $10^{40}-10^{41}$ erg s$^{-1}$ for He 2-10 (Hensler et al. 1997), to $5.7\times10^{38}$ erg s$^{-1}$ for NGC 1569 (Heckman et al. 1995), to $3.5\times10^{39}$ erg s$^{-1}$ for NGC 4449 (Bomans et al. 1997) and to $6.5\times10^{38}$ erg s$^{-1}$ for NGC 5253 (Martin & Kennicutt 1995). Nevertheless, the two superbubbles in NGC 1569 also reach individually such a low X-ray brightness as NGC 1705 in total while in NGC 4449 all the sources are much brighter.

## 3. Discussion

NGC 1705 has unusual properties in the X-rays, UV, and optical. It is therefore a most interesting SBDG. Two mysteries of the X-ray observations require explanation: (1) the two maxima are extremely soft, and (2) this soft radiation is able to escape from NGC 1705 and to penetrate through the MWG, both with individual HI column densities that should each absorb the soft X-ray photons totally.

While we will present an analysis of the hot superbubble and its relationship to the starburst in Paper II, here we wish to draw the attention to the importance of the very soft X-ray spectrum and to discuss its nature and implications.

One possibility is that the two X-ray maxima are produced by Galactic foreground sources with soft spectra. Stellar objects like white dwarfs or nova-like binaries could possibly play this role. For example, a new class of objects discovered with ROSAT, the supersoft X-ray sources (Hasinger 1994), can have very soft spectra. The striking combination of a coincidence of two sources within the two NGC 1705 H$\alpha$ loops (with a 1' separation) and the rarity of soft stellar sources rule out this possibility. In addition, no EUVE source was detected toward NGC 1705, which would be expected for a low-absorption Galactic foreground source of soft X-rays.

Therefore, we prefer the solution that both soft X-ray maxima belong to NGC 1705 for several reasons. This seems possible because we have not yet explored in detail to what degree the Galactic HI column density in direction to NGC 1705 deviates from the Dickey-Lockman value on smaller spatial scales than their survey. More refined radio observations are required to continue the $N_{HI,GAL}$ study in this direction begun by E.M. Arnal (1995). While a quantitative measure of the total $N_{HI}$ cannot be given from the PSPC spectrum above, any value below $10^{20}$ cm$^{-2}$ is probably allowed.

Unfortunately, a problem may also exist within NGC 1705. HL97 mention that Meurer has measured a column density $N_{HI}$ of $2.0\times10^{21}$ cm$^{-2}$ with a 0.5 arcmin beam toward NGC 1705. Such a high column density would totally prevent any soft X-rays from reaching us. On the other hand, most of the gas, more than 90% of the $N_{HI}$, must then belong to NGC 1705. A comparison of the HI distribution of NGC 1705 (Meurer 1994) with the X-ray maxima would require them to be deeply embedded into a HI disk of almost 2' thickness and almost 10' radial extension. This is clearly impossible and therefore we must consider peculiar geometries. In this regard it is noteworthy that HL97 find *"either the SSC NGC 1705-1 is sitting in front of 92% of the HI, or we are seeing through a hole in the HI."* This is consistent with the low reddening of the SSC stars.

Let us now assume as a working hypothesis that we are dealing with a single coherent superbubble that originates from multiple SNII explosions of massive stars in the single central SSC. Accordingly, the two maxima then stem from the upper plumes at both sides of the superbubble which expands symmetrically to the HI disk. In addition, this picture requires that the two still connected "balloons" expand without having swept up sufficient ambient cool gas to form an HI shell. If such a shell has $N_{HI} > 10^{20}$ cm$^{-2}$ it would completely block the soft X-rays. This picture agrees with the presence of H$\alpha$ loops surrounding the X-ray maxima. The ridge between the X-ray plumes, which forms the root of the superbubble, is hidden behind the main H$\alpha$ body.

The presence of two well-defined, almost symmetrical sections of a large superbubble which shine through the HI would seem to require a well organized HI disk within NGC 1705 on a spatial scale of hun-dreds of parsecs. If we further assume a symmetric H I distribution, we would also judge that the SSC, as well as the vertically and symmetrically expanding superbubble, is lying off-center in the disk of NGC 1705 towards us.

Because a narrow distribution of photon energies is required to fit the ROSAT PSPC spectrum and because the Galactic $N_{H I}$ becomes optically thin below $N_{H I,GAL}=10^{20}$ cm$^{-2}$, it is instructive to consider fits to an extreme case where foreground Galactic absorption is only $N_{H I,GAL}=1.0 \times 10^{19}$ cm$^{-2}$. We obtained the best spectral model for this case with the Raymond-Smith of a cooling plasma with a temperature of only around 0.23 keV. The NGC 1705 superbubble is cooler than those detected by ROSAT in most SBDGs.

The star-formation rate is determined from the H$\alpha$ luminosity to 0.21 $M_\odot$ yr$^{-1}$, (Lamb et al. 1985) assuming a Salpeter initial mass function. Accordingly, one finds a SNII rate of 0.0011 yr$^{-1}$, implying a luminosity contribution of $L_{SN} = 3.5 \times 10^{40}$ erg s$^{-1}$, where each SNII supplies $10^{51}$ erg. We can then compare the expected mechanical power input from SNII with the $L_{mech}$ that is needed to drive the observed X-ray-emitting bubble. The radius of the NGC 1705 superbubble is 350-400 pc. In estimating its power requirements, we assume an ambient medium density of 0.4-1.0 cm$^{-3}$ and an expansion age of 12 Myr. An adiabatic solution for the bubble expansion gives the power as $L_{mech} = (1.0 - 4.5) \times 10^{38}$ erg s$^{-1}$ in good agreement with the observed $L_X$. Although the implied SNII power $L_{SN}$ is about five times this value, the maximum power $L_{mech,max}$ is still well below the theoretically predicted supply. Using more sophisticated models, such as those by MacLow & McCray (1988) with conductive heating at the low temperature of kT=0.2 keV seen in NGC 1705, one can obtain agreement between $L_{mech}$, $L_X$, and the temperature of the hot gas. We discuss these issues in detail in Paper II.

From the X-ray observation of NGC 1705 presented here the conclusion are the following: (1) The Galactic H I disk has a hole toward NGC 1705 as the $N_{H I}$ must be very low to transmit the observed very soft X-rays. (2) Similarly, the X-ray sources cannot be embedded in an H I disk within NGC 1705 or they would be absorbed. These two unusual factors could account for the presence in NGC 1705 of the coolest and faintest X-ray source yet observed in a SBDG. Further multi-wavelength observations are also required to complete a solution to the puzzles presented by NGC 1705. Because of its unique location along an unusually clear path within the MWG and its interesting evolutionary state, NGC 1705 provides fundamental insights into the physics of starburst-powered superbubbles and their effects on the structure of its host galaxies.

The authors (G.H., J.S.G.) gratefully acknowledge enlightening discussions with Ginevra Trinchieri and Dominik Bomans. We appreciate the support by M. Arnal with $N_{H I,GAL}$ measurements towards NGC 1705. Our special thanks go to the referee Gerhardt Meurer for his detailed report and constructive comments which helped to clarify the issue of this Paper I. This project has been supported by the Deutsche Agentur für Raumfahrtangelegenheiten (DARA) GmbH under grant no. 50 OR 9302 4 (R.D., N.J.).

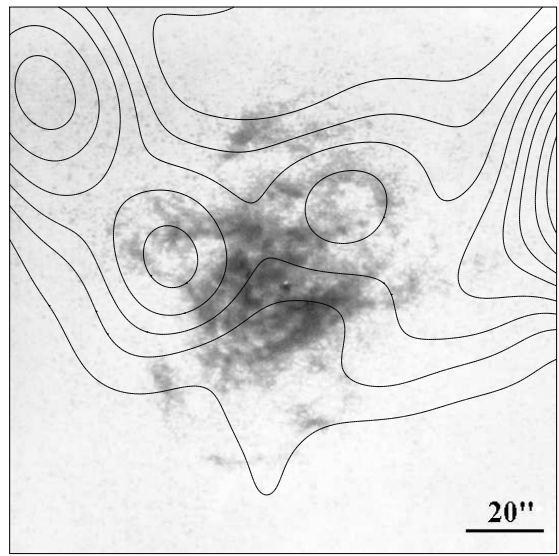

Fig. 1.— Overlay of the ROSAT PSPC contours of NGC 1705 with the Hα image taken from Meurer (1989). The contour levels correspond to 3, 5, 7, 9, 12, 15, 20, and 25 σ where σ is $3.1 \times 10^{-4}$ cts arcmin$^{-2}$ s$^{-1}$. North is up, east to the left. The two brighter sources to the NE and W are likely to be background objects, as discussed in the text.

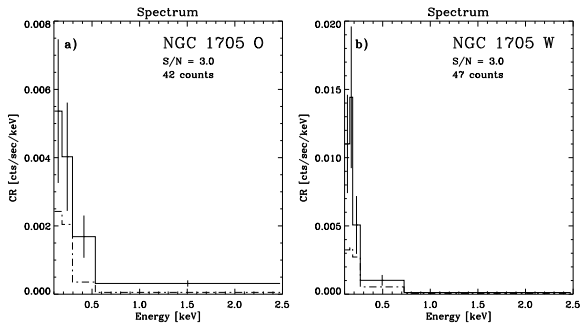

Fig. 2.— Background-subtracted X-ray spectra of both maxima (taken with the ROSAT PSPC) east and west of the main Hα body but enveloped by the Hα loops in NGC 1705 (see fig.1). The solid line joins the measured spectral count rates centered to the maxima; the dash-dotted line represents the background flux outward of the 3 σ contours to either side of a maximum. The integrated X-ray luminosity of NGC 1705 over the Hα body is deduced to $L_X = 1.26 \times 10^{38}$ erg s$^{-1}$.